\documentclass[aps,pra,reprint,noeprint,superscriptaddress,nobibnotes]{revtex4-1}
\pdfoutput=1
\usepackage[utf8]{inputenc}
\usepackage[english]{babel}
\usepackage{microtype}
\usepackage{amsmath,amssymb,amsfonts,mathtools}
\usepackage{braket}
\usepackage{bm}
\usepackage{multirow,booktabs}
\usepackage{graphicx}
\graphicspath{{}}

\usepackage[
    pdftitle={},
    pdfauthor={},
    colorlinks=true,
    unicode=true,
    pdfborder={0 0 0},
    allcolors=blue
]{hyperref}

\usepackage{titlesec}
\titleformat{\subsection}[runin]{\normalfont\bfseries}{\thesubsection}{}{}[.]

\renewcommand{\vec}[1]{\bm{\mathrm #1}}
\newcommand{\vk}{{\vec k}}

\begin{document}

\title{Classification and characterization of nonequilibrium Higgs modes\\
in unconventional superconductors}

\author{L. Schwarz}
\thanks{Both authors contributed equally to this work}
\affiliation{Max Planck Institute for Solid State Research,
70569 Stuttgart, Germany}

\author{B. Fauseweh}
\thanks{Both authors contributed equally to this work}
\affiliation{Max Planck Institute for Solid State Research,
70569 Stuttgart, Germany}

\author{N. Tsuji}
\affiliation{RIKEN Center for Emergent Matter Science (CEMS),
Wako 351-0198, Japan}

\author{N. Cheng}
\affiliation{Department of Physics and Astronomy,
University of British Columbia, Vancouver V6T 1Z1, Canada}

\author{N. Bittner}
\affiliation{Max Planck Institute for Solid State Research,
70569 Stuttgart, Germany}
\affiliation{Department of Physics, University of Fribourg,
1700 Fribourg, Switzerland}

\author{H. Krull}
\affiliation{Lehrstuhl für Theoretische Physik I,
Technische Universität Dortmund, 44221 Dortmund, Germany}

\author{M. Berciu}
\affiliation{Department of Physics and Astronomy,
University of British Columbia, Vancouver V6T 1Z1, Canada}
\affiliation{Quantum Matter Institute, University of British Columbia,
Vancouver V6T 1Z4, Canada}

\author{G. S. Uhrig}
\affiliation{Lehrstuhl für Theoretische Physik I,
Technische Universität Dortmund, 44221 Dortmund, Germany}

\author{A. P. Schnyder}
\affiliation{Max Planck Institute for Solid State Research,
70569 Stuttgart, Germany}

\author{S. Kaiser}
\affiliation{Max Planck Institute for Solid State Research,
70569 Stuttgart, Germany}
\affiliation{4th Physics Institute and Research Center SCoPE,
University of Stuttgart, 70569 Stuttgart, Germany}

\author{D. Manske}
\thanks{Correspondence and requests for materials
should be addressed to D.M. (d.manske@fkf.mpg.de)}
\affiliation{Max Planck Institute for Solid State Research,
70569 Stuttgart, Germany}

\date{\today}

\begin{abstract}
Recent findings of new Higgs modes in unconventional superconductors
require a classification and characterization of the modes
allowed by nontrivial gap symmetry.
Here we develop a theory for a tailored nonequilibrium quantum quench
to excite all possible oscillation symmetries of a superconducting condensate.
We show that both a finite momentum transfer and quench symmetry
allow for an identification of the resulting Higgs oscillations.
These serve as a fingerprint for the ground state gap symmetry.
We provide a classification scheme
of these oscillations and the quench symmetry
based on group theory for the underlying lattice point group.
For characterization, analytic calculations
as well as full scale numeric simulations
of the transient optical response
resulting from an excitation by a realistic laser pulse are performed.
Our classification of Higgs oscillations allows us
to distinguish between different symmetries of the superconducting condensate.
\end{abstract}

\maketitle

\section*{Introduction}
\begin{figure*}[t]
    \centering
    \includegraphics[width=\textwidth]{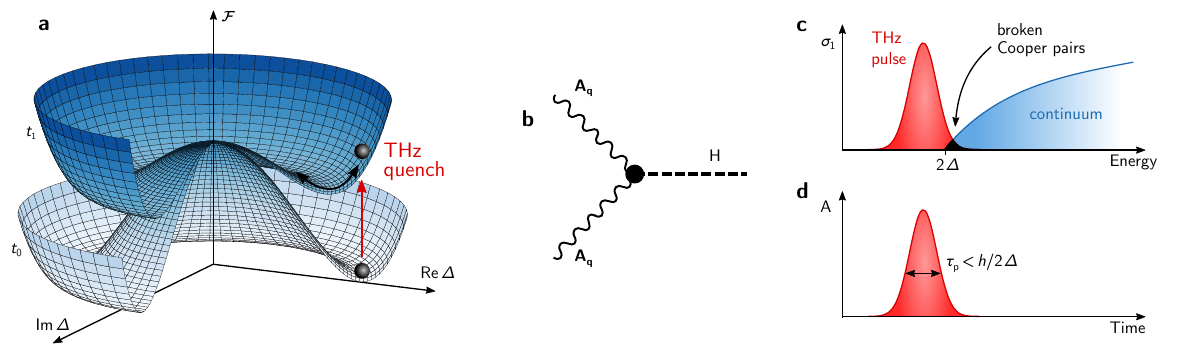}
    \caption{%
    \textbf{Illustration of Higgs oscillations in a superconductor.}
    a) Free energy landscape $\mathcal F$ of a superconductor
    as a function of the real and imaginary part
    of the superconducting gap $\Delta$.
    After a quench at $t_1 > t_0$, the free energy is suddenly changed,
    exciting the superconducting condensate
    and leading to collective Higgs oscillations,
    indicated by a black arrow.
    The red arrow indicates a quench by a THz light pulse.
    b) Feynman diagram describing the excitation
    of a Higgs mode $H$ by a light field $\vec{A}$
    using the Raman vertex.
    An infrared excitation of the Higgs mode (not considered here)
    is only possible if an external current is present.
    c) Higgs excitation mechanism using a THz quench pulse.
    The quench pulse only slightly overlaps
    with the quasi-particle continuum indicated in blue.
    The Mexican hat shrinks due to the breaking of Cooper pairs.
    d) To excite the Higgs oscillation
    the pulse must fulfill the nonadiabaticity condition in time domain.
    }
    \label{fig:figure1}
\end{figure*}
The Higgs mode in superconductors is a collective oscillation
of the order parameter $\Delta$ with the characteristic frequency of $2\Delta$.
It can be understood as a massive excitation
along the radial direction in the Mexican hat potential of the free energy
(see Figure~\ref{fig:figure1}a)
\cite{PhysRevLett.13.508,JLowTempPhys.126.901,PhysRevB.84.174522,%
Pashkinscience14,annurev.varma2015}.
The charge neutral Higgs mode does not couple to linear optical probes
and therefore was expected to be observable
only in materials with competing orders
\cite{PhysRevLett.47.811,PhysRevB.26.4883},
for which it was measured in Raman experiments
\cite{PhysRevB.89.060503,PhysRevB.97.094502,PhysRevLett.122.127001}.
However, an impulsive excitation of Higgs oscillations in nonequilibrium
is possible via a nonlinear process
by quenching the Mexican hat potential with an ultrafast THz light pulse.
Such a quantum quench was demonstrated for the first time
in the $s$-wave superconductor Nb$_{1-x}$Ti$_x$N
\cite{matsunagaPRL12,matsunagaPRL13,Matsunagascience14,PhysRevB.96.020505}.

There are several works indicating that the spectrum of Higgs modes
can be more complex if nontrivial gap symmetries are involved.
Studies on multiband superconductors like MgB$_2$ show
that the Higgs oscillation spectrum contains Higgs modes for both gaps
as well as the Leggett mode, the relative phase mode
\cite{krull_nat_commun_2016,PhysRevB.95.104503}.
Additional Higgs modes with lower energies,
representing oscillations of the gap in different symmetry channels,
were proposed for $d$-wave superconductors
under the assumption of a composite pairing interaction
\cite{PhysRevB.87.054503}.
Besides first quench-probe experiments on cuprates
\cite{mansartPNAS13,PhysRevLett.120.117001,arxiv.1910.07695}
a recent experiment on several types of cuprates
shows clear fingerprints of a $2\Delta$ Higgs mode and a so far unknown
additional mode below $2\Delta$ \cite{arxiv.1901.06675}.
These findings require both a deeper understanding
and a classification and characterization of Higgs modes in nonequilibrium.

So far all descriptions
on how to excite Higgs oscillations with a quench pulse
are working within the dipole approximation,
i.e. neglecting the small wave momentum~$\vec q$ of the external field.
There are other studies which show
that a linear coupling of the vector potential to the condensate
is possible if momentum transfer is involved,
either by impurity scattering in dirty superconductors
\cite{JPhysSocJpn.84.114711,PhysRevB.96.155311,JPhysSocJpn.87.024704,%
PhysRevB.99.224510,PhysRevB.99.224511}
or in current-carrying states
\cite{PhysRevLett.118.047001,PhysRevLett.122.257001}.

Independent of the actual coupling to the external field,
the following instructive picture can be drawn
to understand the excitation process by an ultrashort THz pulse,
where Higgs oscillations are excited
by taking the superconductor out of equilibrium
\cite{Papen07,Papen08,Schny11,Unter08,Krull14,%
PhysRevB.95.104507,zachmannNJP13,Akba13,krull_nat_commun_2016,%
PhysRevB.93.094509,PhysRevB.92.224517,PhysRevB.96.184518}.
Hereby, Cooper pairs are partially broken
and the landscape of the free energy changes suddenly,
i.e. the Mexican hat shrinks.
Thus the THz laser pulse acts like a quantum quench
\cite{volkov1974,barankov2004,Yuz05,Yuz06,YuzJofPhys05,%
peronaci_capone_PRL_15},
reflecting the impulsive character of the light pulse.
As long as this process is faster than the time scale of the condensate,
given by $\tau_\Delta = h/(2 \Delta)$,
where $2 \Delta$ is the energy gap of the superconductor,
the condensate is unable to follow the minimum of free energy adiabatically
\cite{krull_nat_commun_2016,Akba13}.
Consequently, collective Higgs oscillations of the gap are excited,
as sketched in Figure~\ref{fig:figure1}a.

In order to excite Higgs oscillations,
the laser pulse must fulfill two conditions.
On the one hand,
the pulse should only excite a small fraction of the Cooper pairs,
enough to generate a significant quench of the Mexican hat,
but not too many
that the superconducting signatures would be screened by hot electrons.
More specifically, a short optical pulse far above gap frequencies
induces a strong Drude-peak in the optical conductivity,
which would overlap with the weak signal of the Higgs oscillations.
Instead a suitable pulse corresponds to a peak located in or close to the gap,
which only slightly overlaps with the continuum of quasi-particles,
as depicted in Figure~\ref{fig:figure1}c.
On the other hand, the pulse must fulfill the nonadiabaticity condition,
which implies a short laser pulse (Figure~\ref{fig:figure1}d)
and hence requires the broad spectrum in energy domain
(rather than a narrow band multicycle pulse tuned close to the gap).
For typical gaps in the meV regime,
a single cycle THz laser pulse is exactly on the brink of these two regimes
\cite{Fulop12},
allowing for an excitation of the Higgs oscillations
without heating or photo-doping the system too much.

In this article we show how in a nonequilibrium setup for superconductors
with pairing interaction in a single channel, e.g. pure $d$-wave,
oscillations of the condensate in other symmetry channels
can lead to additional Higgs modes as well.
We classify these oscillations of the condensate
based on the irreducible representations
of the point group of the underlying lattice.
The resulting Higgs modes depend on the excitation symmetry
and the ground state symmetry.
Our detailed analysis shows that a full description
of the excitation process requires to go beyond the dipole approximation
in a Raman-like process
and to retain the wave momentum $\vec q$ (see Figure~\ref{fig:figure1}b),
which plays a crucial role in breaking the symmetry of the condensate
and exciting non-$A_{1\mathrm{g}}$ oscillations
of the superconducting condensate.
We show that nonequilibrium Higgs oscillations offer a unique way
to investigate the symmetry and collective excitation spectrum
of superconductors,
which allows to completely characterize the nature
of a superconducting condensate with a single class of experiments.

\section*{Results}
\subsection*{Quantum quenches}
While the nonequilibrium probe of collective excitations
in conventional $s$-wave superconductors has been studied intensively
\cite{Papen07,Papen08,Schny11,Krull14,PhysRevB.95.104507,PhysRevB.92.064508,%
PhysRevB.93.180507,PhysRevB.94.224519,PhysRevB.97.094516},
the response for unconventional superconductors is still in its infancy.
These systems often exhibit very complicated correlations
\cite{DalScience2013,PhysRevX.4.041046}
and a variety of different mechanisms which can lead to superconductivity.
If we want to examine the nonequilibrium response
of the coherent condensate of such superconductors in general,
we have to go back to the fundamental properties of these systems:
The symmetry of the lattice.

According to group theory,
every configuration of the condensate at a given time can always be decomposed
with respect to the different irreducible representations
of the point group symmetry of the lattice.
As an example,
we take a superconductor on a lattice
with $D_{4\mathrm{h}}$ space group symmetry,
which is the lattice symmetry of cuprate high-$T_\mathrm c$ superconductors.
Based on this argument, there are four different irreducible representations:
$A_{1\mathrm{g}}$, $A_{2\mathrm{g}}$, $B_{1\mathrm{g}}$ and $B_{2\mathrm{g}}$.
The condensate oscillations can always be decomposed
into the contributions from these sectors.

\begin{figure}[t]
    \centering
    \includegraphics[width=0.49\textwidth]{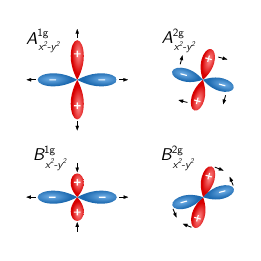}
    \caption{%
    \textbf{Illustration of $d$-wave
    condensate oscillation symmetries.}
    Possible condensate oscillation symmetries
    for a $d_{x^2-y^2}$-wave superconductor
    with point group symmetry $D_{4\mathrm{h}}$ of the underlying lattice.
    The arrows indicate the motion of the lobes as a function of time.
    The notation of the gap symmetry in the subscript
    stresses the initial state,
    from which the
    oscillations of the condensate are excited.
    }
    \label{fig:figure2}
\end{figure}

We start our theoretical description
by considering a quench of the initial state.
Every quantum quench deforms the
condensate from its equilibrium value,
which then can be decomposed into contributions
from different irreducible representations.
Taking a momentum independent quench for example
would only probe the $A_{1\mathrm{g}}$ channel of the condensate
\cite{peronaci_capone_PRL_15},
but does not couple to other allowed symmetries.
The solution to address also the other possible symmetries
is to modify the quench and make it momentum dependent,
so that we can probe other symmetry sectors as well.
For example in case of $d_{x^2-y^2}$-wave superconductivity,
the possible oscillations of the condensate
are shown in Figure~\ref{fig:figure2}.

In order to illustrate this concept,
we perform numerical simulations for $s$- and $d$-wave BCS superconductors
to study the nonequilibrium response to momentum-dependent quantum quenches.
The Hamiltonian we are investigating is given by
\begin{align}
    \label{eq_def_bcs_ham}
    H_{\text{BCS}} = H_0
    - V \sum_{\vk, \vk'} f_\vk f_{\vk'}
    c^{\dagger}_{\vk\uparrow}
    c^{\dagger}_{-\vk\downarrow}
    c^{\ }_{-\vk'\downarrow}
    c^{\ }_{\vk'\uparrow},
\end{align}
with $f_\vk$ describing the symmetry of the interaction,
$V$ the interaction strength, and the normal state Hamiltonian
$H_0 = \sum_{\vk\sigma}\epsilon_{\vk}c^{\dagger}_{\vk\sigma}c^{\phantom{\dagger}}_{\vk\sigma}$,
where $c^{\dagger}_{\vk\sigma}$ creates electrons with momentum $\vk$
and spin $\sigma$.
Within the BCS solution, the gap is determined by
\begin{align}
    \label{gap_eq}
    \Delta_{\vk} &= \Delta f_\vk\,, &
    \Delta &= V \sum_{\vk} f_{\vk}
    \braket{c_{-\vk\downarrow}c_{\vk\uparrow}}.
\end{align}
Now we perform a quantum quench
by changing the symmetry of the condensate
$\braket{c_{-\vk\downarrow}c_{\vk\uparrow}}$
slightly away from its equilibrium value

\begin{align}
    \braket{c_{-\vk\downarrow}c_{\vk\uparrow}}
        = \frac{\Delta f_\vk}{2E_\vk}
    \quad\rightarrow\quad
    \braket{c_{-\vk\downarrow}c_{\vk\uparrow}}'
        = \frac{\Delta f'_\vk}{2E'_\vk}
\end{align}
with $f'_\vk = f_\vk + \delta f^{\mathrm q}_\vk$,
where $f^{\mathrm q}_\vk$ is the quench symmetry
and $\delta$ the quench strength.
The equilibrium value of the condensate has the symmetry
of the gap $f_\vk$ which is not changed and
always remains in a single symmetry sector.
After the quench we calculate the Higgs oscillation
of the order parameter as a function of time
by evaluating the time-dependent gap equation \eqref{gap_eq},
which sums the oscillations of the condensate in momentum space.
For the temporal evolution we use Anderson pseudospin formalism
\cite{PhysRev.112.1900},
where the time-evolution is governed by Bloch equations.
More details are given in the methods.

\begin{figure*}[t]
    \centering
    \includegraphics[width=\textwidth]{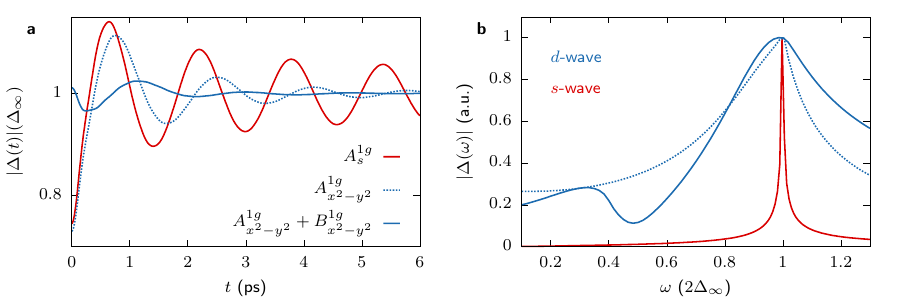}
    \caption{%
    \textbf{Higgs oscillations of a $d$-wave superconductor.}
    a) Numerical simulation of the Higgs oscillations
    induced by various quench symmetries
    for a $d_{x^2-y^2}$-wave superconductor.
    The solid (dotted) blue line shows the gap oscillations
    after a $f^{\mathrm q}_\vk \sim 1$ ($f^{\mathrm q}_\vk \sim x^2-y^2$)
    quench as a function of time.
    The red solid line shows an
    $f^{\mathrm q}_\vk \sim 1$ quench
    for an $s$-wave superconductor for comparison.
    b) Fourier spectrum $|\Delta(\omega)| = |\operatorname{FT} |\Delta(t)||$
    of the Higgs oscillations.
    The  oscillation
    for the $d$-wave gap
    excited by the $f^{\mathrm q}_\vk \sim x^2-y^2$ quench
    shows a single peak,
    similar to the $s$-wave case.
    The peak position corresponds to $2 \Delta_\infty$
    which is two times the maximum of the gap for $t \rightarrow \infty$
    after the quench.
    For the $f^{\mathrm q}_\vk \sim 1$ quench
    a second peak at low energy appears
    resulting from $B^{1\mathrm{g}}_{x^2-y^2}$ oscillations of the condensate.
    }
    \label{fig:figure3}
\end{figure*}

For a given gap symmetry,
there are different oscillations possible for the condensate,
which can be excited depending on the symmetry of the quench.
We use a new notation to describe this oscillation symmetry,
which takes the gap symmetry into account.
We add as an additional information the gap symmetry as subscript
to the group theoretic notation of the irreducible representation name.
We observe that depending on gap and quench symmetries
not only the well known $2\Delta$ Higgs mode occurs
in the spectrum of the Higgs oscillation,
which appears independent on the quench due to coupling between the modes,
but also a second mode at lower energy.

Two examples for this observation are shown in Figure~\ref{fig:figure3}.
In Figure~\ref{fig:figure3}a,
we see the Higgs oscillations of the $d_{x^2-y^2}$ gap
after a $f^{\mathrm q}_\vk \sim x^2-y^2$ and $f^{\mathrm q}_\vk \sim 1$ quench
which excites $A_{x^2-y^2}^{1\mathrm{g}}$ or $B_{x^2-y^2}^{1\mathrm{g}}$
oscillations of the condensate.
The $f^{\mathrm q}_\vk \sim 1$ quench for a $s$-wave superconductor
is shown in red for comparison.
We highlight that the $d$-wave oscillations decay much faster
than the $s$-wave oscillations.
This can be traced back to the stronger dephasing of the mode
due to coupling to the gapless quasi-particles in the $d$-wave case.
The final value of the gap, i.e. $\Delta_{\infty}$,
depends on the strength of the quench,
i.e. how strongly the initial states deviate
from the equilibrium state \cite{Yuz05,Yuz06}.
Figure~\ref{fig:figure3}b
shows the Fourier transform of the Higgs oscillations.
The large peak at $2 \Delta_{\infty}$
corresponds to the symmetric $A_{x^2-y^2}^{1\mathrm{g}}$
oscillation of the condensate.

Most importantly a second low energy mode is visible
for the $d$-wave superconductor
after the $f^{\mathrm q}_\vk \sim 1$ quench.
This mode does not exist for pure $s$-wave superconductors
and it is also not excitable by
the $f^{\mathrm q}_\vk \sim x^2-y^2$ quench in the $d$-wave case,
as the quench has the same symmetry as the ground state gap.
Similarly for other combinations of gap and quench symmetries
additional modes can be identified.
Thus there exists a direct connection between the symmetry of the gap
and the existence of low energy Higgs modes.

To understand the nature of the second mode in more detail,
we perform a linear analysis of the gap dynamics after a quantum quench.
Specifically, we analytically compute the dynamics of the gap
according to the expansion
\begin{align}
    \Delta(t) = \Delta(0) + \delta \Delta(t),
\end{align}
in the first order of $\delta \Delta(t)$ for different initial states.
Here $\Delta(0)$ is the gap at time $t=0$
directly after the quantum quench.
Transforming into Laplace space with complex frequency $s$
allows us to identify the leading contributions
to the gap dynamics and leads to the expression
\begin{align}
    \delta \Delta(s) = \frac{F_2(s)}{1-F_1(s)}
\end{align}
with
\begin{align}
    &F_2(s) \propto \notag\\
    & \int_0^{2\pi} \mathrm{d} \varphi \frac{
        f(\varphi) (\Delta f'(\varphi) - \Delta(0) f(\varphi)
    )
    }{\sqrt{s^2 - 4 (\Delta^2f'(\varphi)^2 - \Delta(0)^2f(\varphi)^2) }
    } \dots
\end{align}
where the dots imply additional weighting factors.
For $F_1(s)$, we find the same denominator in the integrand.
In these expressions we assume
that the symmetry functions primarily depend on the angle $\varphi$
between $\mathbf{k}$ and the $\vk_x$-axis.
For further details on the calculation,
please see Supplementary Note~1.
We can see that the spectrum of Higgs oscillations is controlled
in a nontrivial way by $f(\varphi)^2 - f'(\varphi)^2$
integrated over $\varphi$ weighted by additional factors.
Thus we can trace back the second mode
to the difference in the symmetry between the quench and the condensate.
A second mode in the Higgs oscillation spectrum is only visible
if this difference leads to a second minimum in the denominator.
Particularly, this happens if there are changes of the nodal directions
in the condensate symmetry compared to the equilibrium value.
Due to the nonexistence of nodes in the $s$-wave case,
there will be no second mode visible in the Higgs oscillations
for all possible condensate oscillations.

\subsection*{Realistic pulse}

\begin{figure*}[t]
    \centering
    \includegraphics[width=\textwidth]{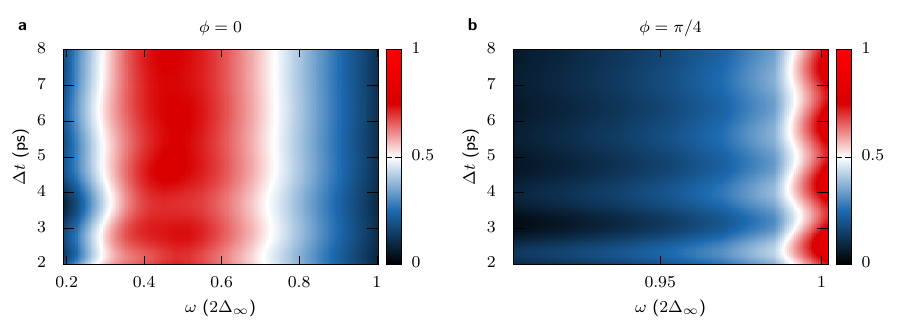}
    \caption{%
    \textbf{Optical conductivity of a $d$-wave superconductor
    after excitation with a quench pulse.}
    Real part of the optical conductivity Re $\sigma(\Delta t, \omega)$
    after a realistic quench pulse for incident angles
    a) $\phi = 0$ and
    b) $\phi = \pi/4$ for a $d_{x^2-y^2}$-wave superconductor.
    The pulse parameters are $\tau_{\textrm{p}} = 0.4$~ps,
    $| {\bf A}_{\textrm{p}} |$ = $7 \cdot 10^{-8}$~Js C$^{-1}$m$^{-1}$
    and $\hbar \omega = 3$~meV for the quench pulse
    and $\tau_{\textrm{p}} = 0.25$~ps,
    $| {\bf A}_{\textrm{p}} |$ = $1 \cdot 10^{-8}$~Js C$^{-1}$m$^{-1}$
    and $\hbar \omega = 2.5$~meV for the probe pulse.
    The gap value in the simulation is $2 \Delta = 2.7 $~meV.
    The vertical axis denotes the time delay between the excitation
    of the system with the quench pulse and the probe pulse.
    The horizontal axis denotes frequency $\omega$.
    The oscillation frequencies in $\Delta t$
    correspond to the frequencies of the Higgs modes
    as shown in Supplementary Figure~3.
    }
    \label{fig:figure4}
\end{figure*}

So far we concentrated our analysis on quantum quenches,
which we classified according to the deformation symmetry
from the equilibrium value.
To show that these results also carry over to more realistic scenarios,
we calculate the response of a $d_{x^2-y^2}$-wave BCS superconductor
coupled to a laser field.
So called pump-probe experiments have been used
to study the excitation and relaxation processes of superconductors
\cite{DalScience2013,leitensdorferPRL10}.
In a pump-probe experiment the pump pulse excites the system
and after a delay time the probe pulse measures various properties
of the transient dynamics of the system.
Varying the delay time,
the temporal evolution of the system after a perturbation can be studied.
As the purpose of the pump pulse in our setup is to quench the system suddenly,
we call this pulse a quench pulse.

The Hamiltonian describing the quench and probe pulses
is given in the methods.
We use the density matrix formalism \cite{Papen07} to calculate the dynamics,
which is exact for the Hamiltonian in \eqref{eq_def_bcs_ham}.
We assume a short and intense THz laser pulse
which excites the condensate in an anisotropic fashion
and the superconducting gap starts to oscillate.
For all pulses we fixed the pulse duration to $\tau_{\mathrm p} = 0.4$~ps.
With this choice we are in the nonadiabatic regime,
where a generation of collective modes is possible.

Further, we varied the direction of the light wave vector $\vec q$
to study the dependence of the optical conductivity on the quench pulse.
Thus we define the angle $\phi$ between $\vec q$
and the $\vk_x$-axis of the superconductor.
The light wave vector is small compared to the Fermi wave vector
$|\vec q| \ll |\vec k_{\mathrm F}|$ such that there is no excitation
of Fulde-Ferrel-Larkin-Ovchinnikov (FFLO) oscillations.
However, it is large enough to break the condensate symmetry
as it couples offdiagonal elements in the quasi-particle distribution
(see Equation~\eqref{eq:coupling_light}).
This is possible due to the Raman-like excitation,
where the photon momentum can be transferred to the condensate.
By choosing the angle of the quench pulse,
different oscillation symmetries can be addressed selectively
(see  Table~\ref{tab.higgs}).

We compare both methods, quantum quench and quench pulse,
in the Supplementary Figure~1.
Besides the time evolution of the gap,
the quench-probe optical conductivity provides an experimental fingerprint
to observe Higgs oscillations as well.
This was explicitly shown in case of $s$-wave symmetry,
i.e. as the oscillation of the conductivity
depending on the delay time \cite{Krull14}.
Thus we use a quench pulse to induce the Higgs oscillations
and a much weaker probe pulse in the same direction
to measure the optical conductivity.

In Figure~\ref{fig:figure4} the real part of the optical conductivity
Re $\sigma(\Delta t, \omega)$ is shown for the angles $\phi = 0$,
along the anti-nodal direction, and $\phi = \pi/4$, along the nodal direction,
as a function of frequency $\omega$ and time delay $\Delta t$.
These angles correspond to the pulse direction with maximum response
in Table~\ref{tab.higgs}.

For $\phi = \pi/4$ the symmetry breaking happens along the diagonal axis
which excites in addition to the symmetric $A^{1\mathrm{g}}_{x^2-y^2}$
also the $B^{2\mathrm{g}}_{x^2-y^2}$ oscillation of the condensate
and most of the weight is located at
the energy of the $2\Delta$ Higgs mode.
There is no low-lying peak visible in the spectrum
as the $B^{2\mathrm{g}}_{x^2-y^2}$
oscillation does not lead to a second mode.
For $\phi=0$ the $A^{1\mathrm{g}}_{x^2-y^2}$
and $B^{1\mathrm{g}}_{x^2-y^2}$ oscillations are excited,
resulting in the low energy Higgs mode and the
$2\Delta$ Higgs mode
and the spectrum is dominated by an in-gap response.
Most importantly the signal oscillates
with respect to the time delay between quench and probe pulse,
reflecting the excitation of Higgs oscillations.
This is demonstrated in Supplementary Figure~3 in more detail.
Thus the angle-resolved quench-probe experiments
should be able to see Higgs oscillations in the optical conductivity,
which can address different oscillation symmetries of the condensate
depending on the incident angle.

\begin{table*}[t]
    \centering
    \setlength{\tabcolsep}{8pt}
    \begin{tabular*}{\textwidth}{lllll}
\toprule
Gap symmetry $f_\vk$ &
Quench symmetry $f^{\mathrm q}_\vk$ &
Pulse direction $\phi$ &
Condensate oscillation $\braket{c_{-\vk\downarrow}c_{\vk\uparrow}}(t)$ &
Higgs modes \\
\toprule
$s$ \hfill
\multirow{4}{*}{\includegraphics[width=1.3cm]{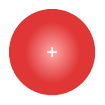}} &
$ 1 $ &
-- &
$A^{1\mathrm{g}}_{s}$ &
\parbox[c]{1em}{\includegraphics[width=0.8cm]{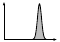}}
\\
&
$ xy \left( x^2 - y^2 \right) $ &
-- &
$A^{2\mathrm{g}}_{s} + A^{1\mathrm{g}}_{s}$ &
\parbox[c]{1em}{\includegraphics[width=0.8cm]{modes1.pdf}}
\\
&
$ x^2 - y^2 $ &
0 &
$B^{1\mathrm{g}}_{s} + A^{1\mathrm{g}}_{s}$ &
\parbox[c]{1em}{\includegraphics[width=0.8cm]{modes1.pdf}}
\\
&
$ xy $ &
$\pi/4$ &
$B^{2\mathrm{g}}_{s} + A^{1\mathrm{g}}_{s}$ &
\parbox[c]{1em}{\includegraphics[width=0.8cm]{modes1.pdf}}
\\
\midrule

$d_{x^2-y^2}$ \hfill
\multirow{4}{*}{\includegraphics[width=1.3cm]{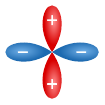}} &
$ x^2 - y^2 $ &
-- &
$A^{1\mathrm{g}}_{x^2-y^2}$ &
\parbox[c]{1em}{\includegraphics[width=0.8cm]{modes1.pdf}}
\\
&
$ xy $ &
-- &
$A^{2\mathrm{g}}_{x^2-y^2} + A^{1\mathrm{g}}_{x^2-y^2}$ &
\parbox[c]{1em}{\includegraphics[width=0.8cm]{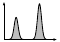}}
\\
&
$ 1 $ &
$0$ &
$B^{1\mathrm{g}}_{x^2-y^2} + A^{1\mathrm{g}}_{x^2-y^2}$ &
\parbox[c]{1em}{\includegraphics[width=0.8cm]{modes2.pdf}}
\\
&
$ xy \left( x^2 - y^2 \right) $ &
$\pi/4$ &
$B^{2\mathrm{g}}_{x^2-y^2} + A^{1\mathrm{g}}_{x^2-y^2}$ &
\parbox[c]{1em}{\includegraphics[width=0.8cm]{modes1.pdf}}
\\
\midrule

$d_{xy}$ \hfill
\multirow{4}{*}{\includegraphics[width=1.3cm]{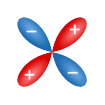}} &
$ xy $ &
-- &
$A^{1\mathrm{g}}_{xy}$ &
\parbox[c]{1em}{\includegraphics[width=0.8cm]{modes1.pdf}}
\\
&
$ x^2 - y^2 $ &
-- &
$A^{2\mathrm{g}}_{xy} + A^{1\mathrm{g}}_{xy}$ &
\parbox[c]{1em}{\includegraphics[width=0.8cm]{modes2.pdf}}
\\
&
$ xy \left( x^2 - y^2 \right) $ &
$0$ &
$B^{1\mathrm{g}}_{xy} + A^{1\mathrm{g}}_{xy}$ &
\parbox[c]{1em}{\includegraphics[width=0.8cm]{modes1.pdf}}
\\
&
$ 1 $ &
$\pi/4$ &
$B^{2\mathrm{g}}_{xy} + A^{1\mathrm{g}}_{xy}$ &
\parbox[c]{1em}{\includegraphics[width=0.8cm]{modes2.pdf}}
\\
\midrule

$g_{xy(x^2-y^2)}$ \hfill
\multirow{4}{*}{\includegraphics[width=1.3cm]{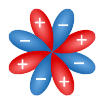}} &
$ xy \left( x^2 - y^2 \right) $ &
-- &
$A^{1\mathrm{g}}_{xy(x^2-y^2)}$ &
\parbox[c]{1em}{\includegraphics[width=0.8cm]{modes1.pdf}}
\\
&
$ 1 $ &
-- &
$A^{2\mathrm{g}}_{xy(x^2-y^2)} + A^{1\mathrm{g}}_{xy(x^2-y^2)}$ &
\parbox[c]{1em}{\includegraphics[width=0.8cm]{modes2.pdf}}
\\
&
$ xy $ &
$0$ &
$B^{1\mathrm{g}}_{xy(x^2-y^2)} + A^{1\mathrm{g}}_{xy(x^2-y^2)}$ &
\parbox[c]{1em}{\includegraphics[width=0.8cm]{modes2.pdf}}
\\
&
$ x^2-y^2 $ &
$\pi/4$ &
$B^{2\mathrm{g}}_{xy(x^2-y^2)} + A^{1\mathrm{g}}_{xy(x^2-y^2)}$ &
\parbox[c]{1em}{\includegraphics[width=0.8cm]{modes2.pdf}}
\\
\bottomrule
    \end{tabular*}
    \caption{%
    \textbf{Classification of Higgs oscillations}
    Possible Higgs oscillations
    for a lattice with $D_{4\mathrm{h}}$ point group symmetry
    shown for $s$, $d_{x^2-y^2}$, $d_{xy}$ and $g_{xy(x^2-y^2)}$
    gap functions (column one).
    A quench can be applied to the condensate
    with a certain symmetry $f^{\mathrm q}_\vk$ (column two),
    which disturbs the ground state condensate symmetry.
    These quenches can be controlled by an incident THz pulse
    with angle $\phi$.
    Pumping at an arbitrary angle corresponds to a quench
    in all symmetry channels.
    Choosing high symmetry direction (column three)
    allows for a selective excitation.
    Such a quench excites oscillations of the condensate (column four),
    classified by the notation of the irreducible representations
    of the lattice symmetry.
    Oscillations of the condensate lead to amplitude oscillations of the gap
    and the qualitative Fourier spectrum of these Higgs oscillations
    is illustrated in the last column, showing the possible Higgs modes.
    An animation on how each quench deforms a given symmetry
    can be found in the Supplementary Movie 1.
    }
    \label{tab.higgs}
\end{table*}

\subsection*{Classification of Higgs oscillations}
Using the results from the quench and quench-probe calculations,
an extension to other symmetries and evaluation of the Higgs oscillations
enables us to write down
a classification scheme for nonequilibrium Higgs modes.
As an example, for the $D_{4\mathrm{h}}$ lattice point group,
the condensate oscillations and resulting Higgs modes
as well as their excitation symmetries are shown in Table~\ref{tab.higgs}
for all fundamental gap symmetries allowed by the point group.

For each gap symmetry $f_\vk$ in column one,
quench symmetries $f^{\mathrm q}_\vk$ in all irreducible representations
of the point group are listed in column two.
Quenching via pump pulses at an arbitrary incident angle $\phi$
results in an excitation of all modes.
Choosing a direction along the high symmetry directions
$\phi = 0$ or $\phi = \pi/4$,
a selective excitation of $B^{1\mathrm{g}}$ or $B^{2\mathrm{g}}$ oscillations
is possible.
These directions,
which correspond to the respective quench symmetry with maximum intensity
are listed in the third column.
As the light pulse always breaks the symmetry
it is in principle not possible
to excite the symmetric $A^{1\mathrm{g}}$ mode alone,
even in the $s$-wave case.
Induced by the quench, the resulting oscillations of the condensate
$\braket{c_{-\vk\downarrow}c_{\vk\uparrow}}$
are shown in column four.
Independent of the quench symmetry,
the symmetric $A^{1\mathrm{g}}$ oscillation is always excited
as any disturbance leads to a global change in the quasi-particle distribution.
The time-dependent amplitude of the energy gap
is calculated from the gap equation \eqref{gap_eq},
where the condensate for each momentum point is summed.
This results in Higgs oscillations of the gap and a schematic picture
of the spectrum is shown in the last column.
For all gap symmetries the $2\Delta$ Higgs mode is visible,
which corresponds to the symmetric $A^{1\mathrm{g}}$
oscillation of the condensate.
Depending on how the non-$A^{1\mathrm{g}}$ oscillations
change the condensate symmetry
from its equilibrium symmetry, i.e. the gap symmetry,
a second Higgs mode is visible in the spectrum.
This is not always the case,
e.g. the $B_{x^2-y^2}^{2\mathrm{g}}$ oscillation is not visible
in the spectrum as a second Higgs mode
despite its asymmetric deviation from the ground state symmetry.
As this oscillation only shifts weight inside the positive and negative lobes
of the $d_{x^2-y^2}$ symmetry but doesn't move the nodal directions,
it will not lead to a second mode
in the summation process for the calculation of the gap oscillations.
Hence, for a full analysis of the gap symmetry
information from multiple quench symmetries are required.
Yet, if we can obtain this information, nonequilibrium Higgs oscillations
can be used as an efficient tool
to completely classify the ground state symmetry of a superconductor.

\section*{Discussion}
To summarize, we introduce
a classification scheme for nonequilibrium Higgs oscillations
which allows to characterize the ground state of superconducting condensates.
Our analytical calculations show
that depending on the symmetry of the quench and of the gap function,
low-lying modes exist,
which can be directly identified
with the different oscillations of the condensate.
We introduce a new notation to combine the information of the ground state
with the quench symmetry
in order to distinguish the different Higgs oscillations.
Simulations of quench-probe experiments
using realistic pulses in a microscopic model show
that the usually ignored wave momentum in the dipole approximation
plays an important role in the excitation
of non-$A_{1\mathrm{g}}$ oscillations of the condensate.
Despite its small value compared to the Fermi wave vector, it is large enough
to break the ground state symmetry and can lead to additional Higgs modes
implementing the proposed analytic quench setup.

It is important to note
that the proposed experimental excitation of the Higgs mode
is a Raman-like excitation and should not be confused with an
infrared-active excitation
\cite{PhysRevLett.122.257001}.
In the latter case, a driven ac current would occur
and thus the strength and polarization of the electric field
is more important than the small momentum of the photon.
In the former case,
the polarization of the electric field plays a minor role
and the photon momentum becomes much more important.

We find that the Higgs modes are visible in the optical conductivity
of the proposed quench-probe experiment,
paving the way for investigations and classifications
of the dynamics of known and unknown superconductors
directly within this framework.
This analysis is applicable for all superconductors
and requires only the knowledge of the symmetry of the crystal.
It is a natural extension of the group theoretical notation
to the case of nonequilibrium excitation of the system.
To demonstrate the approach,
we fully characterize all possible Higgs oscillations
for the important $D_{4\mathrm{h}}$ point group,
relevant for example for high-temperature cuprate superconductors.
This main result is summarized in Table~\ref{tab.higgs},
which goes beyond a simple product table
of gap symmetry and light pulse direction.
The number of excited fundamental condensate oscillations
does not directly correlate with the number of observed Higgs modes,
which depend in a nontrivial way
on the phase and nodal structure of the order parameter.

The experimental realization of the proposed momentum transfer
to break the condensate symmetry will be challenging.
If light couples to the Higgs mode only indirectly via electrons,
momentum scattering on timescales faster than the oscillation period might
wash out or destroy the preferred direction of the pulse.
Depending on the strength of this momentum distribution effect,
the second Higgs mode could be damped,
might no longer follow its predicted angular dependency
or may be even completely suppressed.
On the other hand, recent studies have shown
that impurity scattering in dirty superconductors
even enhances the coupling of light to the condensate
and the excitation of the Higgs mode
\cite{JPhysSocJpn.84.114711,PhysRevB.96.155311,JPhysSocJpn.87.024704,%
PhysRevB.99.224510,PhysRevB.99.224511}.
As no experiments exist so far
which allow to measure the transferred photon momentum in detail,
the field is open for further experimental and theoretical investigations.
However, the current efforts to measure Higgs oscillations
on different cuprates \cite{mansartPNAS13,PhysRevLett.120.117001,%
arxiv.1901.06675,arxiv.1910.07695}
show already fingerprints of collective oscillations.

It is important to note that we do not introduce additional energy scales
nor other degrees of freedom, such as subdominant channels:
the observed oscillations and the corresponding frequencies
are intrinsic to the pure $d$-wave superconductor
and do not require composite pairing symmetries.
Note that we assume that no competing order,
such as a charge density wave, exists.
Otherwise, the spectroscopic signatures of the Higgs oscillations
could be modified due to the interplay between the two phases
\cite{PhysRevB.92.184511}.
Furthermore, effects which can modify the Higgs spectrum as well
are superconductors in the strongly coupled regime \cite{PhysRevB.93.094509},
coupling to Leggett modes in multiband systems \cite{krull_nat_commun_2016}
or collective excitations of pair states in subleading channels,
i.e. an excitation of the Bardasis-Schrieffer mode
\cite{PhysRev.121.1050,PhysRevLett.110.187002,NPJ.Quantum.Mater.3.48,%
PhysRevB.100.140501}.
Other details of the normal state, such as Fermi arcs,
play an unimportant role
after a quantum quench of the superconducting condensate,
as they only modify the scattering processes of the broken Cooper pairs.
This could potentially change the damping of the Higgs oscillations,
but has no effect on our proposed classification scheme.

There are different possibilities
how a symmetry breaking momentum transfer to the condensate
is realized in an experiment.
Tilting the quench pulse direction towards the superconducting plane
induces a finite in-plane momentum of the photons.
More controlled momentum dependent excitations and probes are possible
in a THz-four-wave mixing
\cite{PhysRevLett.109.147403} or transient grating
\cite{Science.300.1410,JChemPhys.120.4755} setup.
Other possibilities include momentum-dependent scattering processes
as well as coupling to other finite-momentum modes.
This is discussed for example for superconductors under external current
\cite{PhysRevLett.118.047001,PhysRevLett.122.257001}
or as a possibility for phonon-coupled amplitudon dynamics
in excitonic insulators
\cite{SciAdv.4.eaap8652,10.1117/12.2304795}.

The classification and characterization of Higgs oscillations
open the possibility to perform spectroscopic studies on superconductors
to determine the symmetry of the order parameter.
Compared to other types of measurements like ARPES
or interferometry experiments using Josephson junctions,
which can retrieve either amplitude or phase information,
spectroscopy of Higgs oscillations with phase-stable THz lasers
allows to determine amplitude and phase
within a single type of quench-probe experiment.
In principle our theory for Higgs spectroscopy
is not limited to characterizing equilibrium condensates
but may also be used to
investigate possible light induced superconducting states
in transient states of matter
\cite{Mitrano2016,Fausti189,JPhysSocJn.88.044704,Kaiser_Scripta_2017}.
Beyond that, Higgs spectroscopy could also be extended
to investigate collective excitations of non-superconducting,
symmetry broken phases,
such as order parameter oscillations in excitonic insulators
\cite{SciAdv.4.eaap8652,10.1117/12.2304795}
or Higgs modes in antiferromagnets \cite{PhysRevLett.119.067201}.

\section*{Methods}

\subsection*{Extended BCS model}
The Hamiltonian we are investigating is given by
\begin{align}
    \label{eq_def_bcs_ham_supp}
    H_{\text{BCS}} &= H_0
        - \sum_{\mathclap{\vk, \vk' \in \mathcal{W}}} V_{\vk\vk'}
        c^{\dagger}_{\vk\uparrow}  c^{\dagger}_{-\vk\downarrow}
        c^{\ }_{-\vk'\downarrow} c^{\ }_{\vk'\uparrow} \,, \\
    H_0 &= \sum_{\vk,\sigma} \epsilon_{\vk}
        c^{\dagger}_{\vk\sigma}c^{\phantom{\dagger}}_{\vk\sigma}.
\end{align}
The normal state Hamiltonian $H_0$ is taken to be a free electron gas
with an effective mass $m$.
The pairing interaction $V_{\vk\vk'} = Vf_\vk f_{\vk'}$
is assumed to be separable with the interaction strength $V$.
The energy dispersion
$\epsilon_{\vk} = \hbar^2\vk^2/(2m)  - \epsilon_{\mathrm F}$
is measured relative to the Fermi level $\epsilon_{\mathrm F}$.
We apply the BCS solution in order to describe the superconducting phase.
The superconducting gap equation reads
\begin{align}
    \Delta_\vk &= \Delta f_\vk\,, &
    \Delta &= V \sum_{\vk\in \mathcal{W} } f_{\vk}
    \braket{c^{\phantom{\dagger}}_{-\vk\downarrow}
        c^{\phantom{\dagger}}_{\vk\uparrow}}\,.
    \label{gap_eq_supp}
\end{align}
The sums in Equations~\eqref{eq_def_bcs_ham_supp} and \eqref{gap_eq_supp}
are taken over the set $\mathcal{W}$ of all $\vk$ vectors
with $|\epsilon_\vk| \leq \hbar \omega_\mathrm{c}$,
$\omega_\mathrm{c}$ being the frequency cutoff.
For a phononic glue, this corresponds to the Debye frequency.
The function $f_\vk$ is the gap symmetry function,
where in the case of an $s$-wave superconductor $f_\vk = 1$ is a constant.
In general, the symmetry function can be decomposed
into the basis functions of the irreducible representations
of the point group of the underlying lattice.
In case of the $D_{4\mathrm{h}}$ group,
the basis functions are shown in Supplementary Table~1,
where the $d$-wave symmetry belongs to the $B_{1\mathrm{g}}$ representation.
For all of our calculations we assume
that there is only a polar angle dependency $\varphi$ on the momentum $\vk$
in the vicinity of the Fermi energy.
Therefore we use the functions shown in the third column.

In the ground state
the expectation values for the electron and quasi-particle distribution read
\begin{align}
    \label{eq.distributions}
    \Braket{c^{\dagger}_{\vk\uparrow}c^{\phantom{\dagger}}_{\vk\uparrow}}
        &= \frac 1 2 - \frac{\epsilon_{\vk}}{2 E_\vk}\,, &
    \Braket{c^{\phantom{\dagger}}_{-\vk\downarrow}
        c^{\phantom{\dagger}}_{\vk\uparrow} }
        &= \frac{\Delta f_\vk}{2 E_\vk}\,,
\end{align}
where $E_\vk = \sqrt{ \epsilon^2_\vk + | \Delta_{\vk} |^2 }$
is the quasi-particle energy.

\subsection*{Anderson pseudospin description}
\label{sec:anders_pseudo}
We define the Nambu-Gorkov spinor
\begin{align}
    \Psi_{\vec{k}} = \begin{pmatrix}
        c_{\vk\uparrow} \\
        c^\dagger_{-\vk\downarrow}
    \end{pmatrix}
\end{align}
and Anderson pseudospin \cite{PhysRev.112.1900}
\begin{align}
    \vec{\sigma}_\vk &=
        \frac{1}{2} \Psi_\vk^\dagger \vec{\tau} \Psi_\vk\,,
\end{align}
where $\vec{\tau}$ are the Pauli matrices.
The BCS Hamiltonian takes the form
\begin{align}
    H_\mathrm{BCS} &= \sum_\vk \vec{b}_\vk \vec{\sigma}_\vk
\end{align}
with
\begin{align}
    \vec{b}_\vk &= \left( -2\Delta f_\vk , 0 , 2\epsilon_\vk\right)\,,
\end{align}
where we assume a fixed phase of the gap such that $\Delta \in \mathbb{R}$.
In equilibrium, the $y$-component of the pseudospin is zero
$\braket{\sigma_\vk^y} = 0$, whereas the $x$- and $z$-component read
\begin{align}
    \braket{\sigma_\vk^x} &= \frac{\Delta f_\vk}{2E_\vk}\,, &
    \braket{\sigma_\vk^z} &= -\frac{\epsilon_\vk}{2E_\vk}\,.
\end{align}
At $t=0$,
we apply a state quench where we change the symmetry of the condensate
by changing the pseudospin expectation values
\begin{align}
    \braket{\sigma_\vk^x}(0) &= \frac{\Delta f'_\vk}{2E'_\vk}\,, &
    \braket{\sigma_\vk^z}(0) &= -\frac{\epsilon_\vk}{2E'_\vk},
\end{align}
where $E'_\vk = \sqrt{ \epsilon_\vk^2  + (\Delta f'_\vk)^2 }$
and $f'_\vk = f_\vk + \delta f^{\mathrm q}_\vk$
with the quench symmetry $f^{\mathrm q}_\vk$ and strength $\delta$.
This changes the initial ground state symmetry of the condensate,
which is the same as the gap symmetry,
to the quenched symmetry $f'_\vk$.
Note that the gap $\Delta(0)$ at $t=0$
is different from the equilibrium gap $\Delta$
due to the sudden change of the system state.
For arbitrary times, the gap equation reads
\begin{align}
    \Delta(t) &= V \sum_\vk f_\vk \braket{\sigma_\vk^x}(t)\,.
    \label{eq:gapt}
\end{align}
This renders the Hamiltonian time-dependent as the pseudomagnetic field
depends on the gap.
The time-evolution of the pseudospin in the quenched system
is described by Bloch equations
\cite{PhysRevB.92.064508}
\begin{align}
    \partial_t \vec{\sigma}_\vk(t) &= \mathrm{i}
        \left[H_\mathrm{BCS}(t), \vec\sigma_\vk(t) \right]
        = \vec{b}_\vk(t) \times \vec{\sigma}_\vk(t)\,.
    \label{eq:bloch}
\end{align}
The Bloch equations \eqref{eq:bloch} can then be solved
together with the time-dependent
gap equation \eqref{eq:gapt} self-consistently.

\subsection*{Coupling to vector potential}
The Hamiltonian describing the coupling
between superconductor and quench pulse,
which brings the system out of equilibrium, is modeled by
\begin{align}
    \label{eq:coupling_light}
    H_{\text{Laser}} &= \frac{e\hbar}{2m}
        \sum_{\vk,\vec q,\sigma}
            (2\vk+\vec q)\vec A_{\vec q}(t)
            c^{\dagger}_{\vk+\vec q,\sigma}
            c^{\phantom{\dagger}}_{\vk,\sigma}\\  \nonumber
    &+ \frac{e^2}{2m}\sum_{\vk,\vec q,\sigma}\left(
        \sum_{\vec q^{\prime}}
        \vec A_{\vec q-\vec q^{\prime}}(t)
        \vec A_{\vec q^{\prime}}(t)\right)
        c^{\dagger}_{\vk+\vec q,\sigma}c^{\phantom{\dagger}}_{\vk,\sigma} \,,
\end{align}
where ${\vec A}_{\vec q}(t)$ is the transverse vector potential
\cite{Papen07, Krull14}.
Working within the Coulomb gauge,
the quench pulse is expressed in terms of the transverse vector potential
\begin{align}
    \label{eq.pulse_equation}
    \vec A_{\vec q}(t)
        &= \vec A_{\textrm{p}} e^{
        - \left( \frac{ 2 \sqrt{ \ln 2} t}{\tau_{\textrm{p}}} \right)^2
    }
    \left(
        \delta_{ {\vec q}, {\vec q}_p } e^{ - i \omega_{\textrm{p}}  t}
        + \delta_{ {\vec q}, - {\vec q}_p } e^{ + i \omega_{\textrm{p}} t}
    \right) .
\end{align}
The quench pulse is of Gaussian shape
with photon frequency $\omega_{\textrm{p}}$,
photon wave vector ${\vec q}_{\textrm{p}}$,
full width at half maximum (FWHM) $\tau_{\textrm{p}}$
and amplitude ${\vec A}_{\textrm{p}}$.
For our simulations we consider various directions of the photon wave vector
${\vec q}_{\textrm{p}}$ and with this concomitantly
various directions of the quench induced by the pulse.

\subsection*{Optical conductivity}
To calculate the optical conductivity,
we calculate the temporal evolution of the current density
as function of the time delay between the quench and probe pulse
\begin{align}
    \vec j_{\vec q_{\mathrm{pr}}} (\Delta t, t) &=
        \frac{-e\hbar}{2mV} \sum\limits_{ \vk, \sigma}
        \left( 2 \vk + \vec q_{\mathrm{pr}} \right)
        \Braket{ c^\dagger_{\vk, \sigma}
        c^{\phantom{\dagger}}_{\vk + \vec q_{\mathrm{pr}}, \sigma}}(\Delta t,t)
        \nonumber \\
    &- \frac{e^2}{m V} \sum\limits_{ \vk, \vec q,  \sigma}
        \vec A_{\vec q_{\mathrm{pr}} - \vec q}
        \Braket{c^\dagger_{\vk, \sigma}
            c^{\phantom{\dagger}}_{\vk + \vec q, \sigma} } (\Delta t, t) ,
\end{align}
where $V$ is the normalization volume and $\vec q_{\mathrm{pr}}$
is the wave-vector of the probe pulse \cite{Papen07}.
We neglect the second term,
because it only leads to an offset of the imaginary part
of the optical conductivity.
Then, the optical conductivity can be calculated
\cite{Krull14} by computing
\begin{align}
    \sigma(\Delta t, \omega) &=
        \frac{j_{\vec q_{\mathrm{pr}}} (\Delta t, \omega)}
        {i \omega A_{\vec q_{\mathrm{pr}}}(\omega)} .
\end{align}

\subsection*{Density matrix formalism}
In order to simulate the evolution of the system,
we use methods based on an expansion of Heisenberg's equation of motion.
For the temporal evolution of the order parameter
we use the density matrix formalism \cite{RevModPhys.74.895}.
The main task of this technique is to derive equations of motion
for quasi-particle densities.
Within this formalism it is advantageous to perform a Bogoliubov transformation
of the electron operators which diagonalizes the Hamiltonian $H_{\text{BCS}}$
in the initial state.
We introduce new fermionic operators $\alpha_\vk $ and $\beta_\vk $, with
\begin{align}
    \alpha_\vk &=
        u_\vk c_{\vk\uparrow} - v_\vk c^{\dag}_{- \vk\downarrow} ,
    \notag\\
    \beta_\vk^{\dag} &=
        v_\vk^* c_{\vk\uparrow} + u_\vk^* c^{\dag}_{- \vk \downarrow} ,
\end{align}
where $u_\vk = \sqrt{ ( 1 + \epsilon_\vk / E_\vk ) / 2} $
and $v_\vk =  \sqrt{ ( 1 - \epsilon_\vk / E_\vk ) / 2} $.
We emphasize that the coefficients $u_\vk$ and $v_\vk$
do not depend explicitly on time, i.e.,
the temporal evolution of the quasi-particle densities
is computed with respect to a fixed time-independent Bogoliubov-de Gennes basis
in which the initial state is diagonal.

All physical observables, such as the order parameter amplitude
$|\Delta_\vk(t)|$ can now be expressed
in terms of the new Bogoliubov quasi-particle densities
$\braket{ \alpha^{\dag}_\vk \alpha^{\ }_{\vk'}}$,
$\braket{ \beta^{\dag}_\vk \beta^{\ }_{\vk'}}$,
$\braket{ \alpha^{\dag}_\vk \beta^{\dag }_{\vk'}}$,
and
$\braket{\alpha^{\ }_\vk \beta^{\ }_{\vk'}}$.
Applying the density matrix formalism for these quasi-particle densities,
we get a closed set of differential equations.
The ensuing differential equations are then solved
on a finite size grid in momentum space.
More details about the implementation can be found in \cite{Papen07,Krull14}.

The pulse solution for a pumping angle of $\phi = 0$
is shown in Supplementary Figure~1.
It shows a stronger broadening
than the analytical calculations, but is in qualitative agreement.
Note that we do not expect quantitative agreement,
since a quantum quench is different from a laser pulse.

\subsection*{Numerical implementation}
\label{sec.numerics}
In our simulations we use the parameters $\Delta = 1.35$~meV,
$E_{\mathrm F}=9470$~meV and $m=1.9$~$m_{\mathrm e}$,
which are motivated by the parameters for lead \cite{Papen07}.
However all our computations can be rescaled to any energy scale for the gap.
The numerical equations are computed on a finite size grid in momentum space
in 2 dimensions similar to \cite{Papen07, Krull14}.
To obtain the required accuracy
to resolve the small wave momentum $\vec q_{\mathrm p}$,
we restrict our grid in a small region around the Fermi energy
with a cutoff of $E_{\mathrm c} = 8.3$~meV.
We have ensured by varying the cutoff energy
that our results do not depend on the discretization range.
The $x$-direction is discretized
with a step size of the wave momentum $\vec q_{\mathrm p}$,
which results in 1000 - 2000 points.
This is advantageous as we can resolve directly the coupling
between the offdiagonal elements like $\braket{\alpha_\vk\beta_{\vk+\vec q}}$.
As the coupling in $y$-direction is only indirect
via the energy gap and therefore much smaller,
we choose between 100 and 500 points for this direction.
In total we have of the order of $10^6$ grid points.
To reduce computational effort,
we consider offdiagonal elements like $\braket{\alpha_\vk\beta_{\vk+n\vec q}}$
only up to $n = 4$ as larger offdiagonal elements only contribute in order
$\mathcal{O} \left( {\vec A}_{\textrm{p}}^5 \right)$.

\section*{Data Availability}
All relevant numerical data are available from the corresponding author
upon reasonable request.

\section*{Code Availability}
The numerical code used to calculate the results for this work
is available from the corresponding author upon reasonable request.

\section*{Acknowledgements}
We thank R. Shimano, M.J. Kim, and H. Chu for fruitful discussions.
We further thank the Max Planck-UBC-UTokyo Center for Quantum Materials
for fruitful collaborations and financial support.
N.T. is supported by JSPS KAKENHI (Grant No. 16K17729)
and JST PRESTO (Grant No. JPMJPR16N7).
N.B. is supported by the ERC Consolidator Grant 724103.
S.K. acknowledges support from the Ministerium für Wissenschaft,
Forschung und Kunst Baden-Württemberg through the Juniorprofessuren-Programm
and a fellowship from the Daimler und Benz Stiftung.
G.S.U. acknowledges funding by the DFG in grant UH 90/13-1.

\section*{Author Contributions}
The numerical calculations have been done by B.F. and L.S.
on the computer cluster of the Max-Planck-Institut für Festkörperforschung.
Initial calculations have been performed by N.C, N.B and H.K.
L.S. wrote the paper together with A.S., S.K. and D.M.
An earlier version was written by B.F. and L.S.
N.T, M.B. and G.S.U contributed
to the discussion and interpretation of the results.

\section*{Competing Interests}
The authors declare no competing interests.

\end{document}


\title{Classification and characterization
of nonequilibrium Higgs modes\\ in unconventional superconductors:\\
Supplementary Information}

\author{L. Schwarz}
\thanks{Both authors contributed equally to this work}
\affiliation{Max Planck Institute for Solid State Research,
70569 Stuttgart, Germany}

\author{B. Fauseweh}
\thanks{Both authors contributed equally to this work}
\affiliation{Max Planck Institute for Solid State Research,
70569 Stuttgart, Germany}

\author{N. Tsuji}
\affiliation{RIKEN Center for Emergent Matter Science (CEMS),
Wako 351-0198, Japan}

\author{N. Cheng}
\affiliation{Department of Physics and Astronomy,
University of British Columbia, Vancouver V6T 1Z1, Canada}

\author{N. Bittner}
\affiliation{Max Planck Institute for Solid State Research,
70569 Stuttgart, Germany}
\affiliation{Department of Physics, University of Fribourg,
1700 Fribourg, Switzerland}

\author{H. Krull}
\affiliation{Lehrstuhl für Theoretische Physik I,
Technische Universität Dortmund, 44221 Dortmund, Germany}

\author{M. Berciu}
\affiliation{Department of Physics and Astronomy,
University of British Columbia, Vancouver V6T 1Z1, Canada}
\affiliation{Quantum Matter Institute, University of British Columbia,
Vancouver V6T 1Z4, Canada}

\author{G. S. Uhrig}
\affiliation{Lehrstuhl für Theoretische Physik I,
Technische Universität Dortmund, 44221 Dortmund, Germany}

\author{A. P. Schnyder}
\affiliation{Max Planck Institute for Solid State Research,
70569 Stuttgart, Germany}

\author{S. Kaiser}
\affiliation{Max Planck Institute for Solid State Research,
70569 Stuttgart, Germany}
\affiliation{4th Physics Institute and Research Center SCoPE,
University of Stuttgart, 70569 Stuttgart, Germany}

\author{D. Manske}
\thanks{Correspondence and requests for materials
should be addressed to D.M. (d.manske@fkf.mpg.de)}
\affiliation{Max Planck Institute for Solid State Research,
70569 Stuttgart, Germany}

\date{\today}

\maketitle

\section*{Supplementary Note 1: Linear analysis}
In order to gain an analytical insight
into the evolution of the nonequilibrium equations of motion,
we consider small quenches $\delta \ll 1$,
such that we can linearize the deviations
\begin{align}
    \braket{\vec \sigma_\vk}(t)
        &= \braket{\vec \sigma_\vk}(0)
            + \braket{\delta \vec \sigma_\vk}(t)\,, \\
    \Delta(t) &= \Delta(0) + \delta \Delta (t)\,.
\end{align}
The linearized Bloch equations read
\begin{align}
    \partial_t \braket{\delta \sigma_\vk^x} (t)
        &= -2 \epsilon_\vk \braket{\delta \sigma_\vk^y}(t)\,, \\
    \partial_t \braket{\delta \sigma_\vk^y} (t)
        &= 2 \epsilon_\vk
            \frac{\Delta f'_\vk-\Delta(0)f_\vk}{2E'_\vk}
        + 2 \epsilon_\vk\braket{\delta \sigma_\vk^x}(t)
        + 2 \Delta(0) f_\vk \braket{\delta \sigma_\vk^z}(t)
        - 2 \epsilon_\vk \frac{f_\vk}{2E'_\vk}\delta \Delta (t)\,, \\
    \partial_t \braket{\delta \sigma_\vk^z} (t)
        &= - 2 f_\vk \Delta(0) \braket{\delta \sigma_\vk^y}(t)\,.
\end{align}
Going into Laplace space with complex frequency $s$,
we obtain algebraic equations and can solve for the gap
\begin{align}
    \delta \Delta(s) &= \frac{F_2(s)}{1-F_1(s)}\,,
\end{align}
where
\begin{align}
    F_1(s) &= C
        - V \sum_\vk f_\vk^2
        \frac{s^2 + 4 \Delta(0)^2 f_\vk^2}
        {2E'_\vk \left(
            s^2 + 4 \epsilon_\vk^2 + 4 \Delta(0)^2 f_\vk^2
        \right)} \,,\\
    F_2(s) &= \frac{\Delta(0)}{s}\left(C-1\right)
        + \frac{1}{s}
        V \sum_\vk
            f_\vk (\Delta f'_\vk - \Delta(0) f_\vk )
            \frac{s^2 + 4\Delta(0)^2 f_\vk^2}
                {2E'_\vk(s^2+4\epsilon_\vk^2 + 4\Delta(0)^2 f_\vk^2)}\,.
\end{align}
with $C = V\sum_\vk \frac{f_\vk^2}{2E'_\vk} \in \mathcal{O}(1)$.
With $U= V D(\epsilon_{\mathrm F})$,
where $D(\epsilon_{\mathrm F})$ is the density of states at the Fermi level,
we replace the momentum sum with an integral over $\epsilon$ and $\varphi$
\begin{align}
    V\sum_\vk \rightarrow U \int_{-\infty}^\infty \mathrm d\epsilon
        \int_0^{2\pi}\mathrm d\varphi\,.
\end{align}
We focus first on the nominator and evaluate the $\epsilon$ integral
\begin{align}
    F_2(s) &= \frac{\Delta(0)}{s}(C-1) + \frac{U}{s}
        \int_{-\infty}^\infty \mathrm{d} \epsilon
            \int_0^{2\pi} \mathrm{d} \varphi\,
        \frac{f(\varphi) (\Delta f'(\varphi) - \Delta(0) f(\varphi) )}
            {2\sqrt{\epsilon^2 + \Delta^2 f'(\varphi)^2 }}
        \frac{s^2 + 4\Delta(0)^2 f(\varphi)^2}
            {s^2+4\epsilon^2 + 4\Delta(0)^2 f(\varphi)^2} \nonumber \\
        &=  \frac{\Delta(0)}{s}(C-1) + \frac{U}{s}
            \int_0^{2\pi} \mathrm{d} \varphi
            \frac{f(\varphi) (\Delta f'(\varphi) - \Delta(0) f(\varphi) )
            \sqrt{s^2 + 4\Delta(0)^2 f(\varphi)^2}}
        {\sqrt{s^2 - 4(\Delta^2f'(\varphi)^2 - \Delta(0)^2 f(\varphi)^2)}}
        \notag\\&\qquad
        \times \tanh^{-1}\left(
            \frac{\sqrt{s^2
                - 4(\Delta^2f'(\varphi)^2 - \Delta(0)^2 f(\varphi)^2)}}
            {\sqrt{s^2 + 4\Delta(0)^2f(\varphi)^2}}
            \right)\,. \label{eq.result_lin_ana}
\end{align}
The expression for $F_1(s)$ reads
\begin{align}
    F_1(s) &= C - U
    \int_{-\infty}^\infty \mathrm{d} \epsilon \int_0^{2\pi} \mathrm{d} \varphi\,
        f(\varphi)^2 \frac{
            s^2 + 4\Delta(0)^2f(\varphi)^2}
        {2\sqrt{\epsilon^2 + \Delta^2f'(\varphi)^2}
            (s^2 + 4\epsilon^2 + 4\Delta(0)^2f(\varphi)^2)}\notag\\
        &= C - U \int_0^{2\pi} \mathrm{d} \varphi\,
        f(\varphi)^2 \frac{
            \sqrt{s^2 + 4\Delta(0)^2f(\varphi)^2}}
        {\sqrt{s^2 - 4(\Delta^2f'(\varphi)^2 - \Delta(0)^2 f(\varphi)^2)}}
        \tanh^{-1}\left(
            \frac{\sqrt{s^2 - 4(\Delta^2f'(\varphi)^2 - \Delta(0)^2 f(\varphi)^2)}}
            {\sqrt{s^2 + 4\Delta(0)^2f(\varphi)^2}}
            \right)\,.
\end{align}
Without explicitly solving the integral over $\varphi$,
we already see that possible modes, i.e. maxima in the function $F_2(s)$,
are controlled by the difference $f(\varphi)^2 - f'(\varphi)^2$.
This expression in the denominator determines the main oscillation frequencies
in combination with the other weighting factors in the integrand.
Depending on the gap and the quench symmetry, a low energy peak may appear
as observed in the main text.

\section*{Supplementary Note 2:
Comparison between analytical calculations and numerical results}
Since we cannot solve the integral \eqref{eq.result_lin_ana} analytically,
we calculate the gap dynamics of the linearized equations numerically
and compare it to the solution of the full equation
in Supplementary Figure~\ref{fig:sfigure1}.
We assume that the order parameter has $d$-wave symmetry,
i.e. $f(\varphi) = \cos(2\varphi)$
according to Supplementary Table~\ref{tab.basis_functions},
and we quench in the $A_{1\mathrm{g}}$ channel, i.e. $f^{\mathrm q}(\varphi) = 1$.
Both solutions show a low energy peak,
however the full nonlinear equations
further amplify the intensity of the low energy mode.
The energy of the lower peak depends on the quench strength,
as a stronger quench increases the difference
between the ground state symmetry $f(\varphi)$
and the quenched symmetry $f'(\varphi)$.

\begin{figure}[t]
\begin{minipage}[t]{0.45\textwidth}
    \centering
    \includegraphics[width=\textwidth]{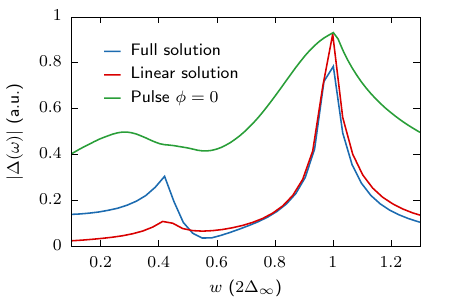}
    \caption{%
    \textbf{Comparison between different solutions for $d$-wave Higgs modes}
    Fourier spectrum $|\Delta(\omega)| = |\operatorname{FT} |\Delta(t)||$ of the gap oscillations induced by a quench $f^{\mathrm q}(\varphi) = 1$ and a quench pulse.
    The dynamics in the quenched case is calculated
    once in the linear approximation and once with the full solution.
    All three results show the $2\Delta$ Higgs mode and a second mode below.
    }
    \label{fig:sfigure1}
\end{minipage}\hfill
\begin{minipage}[t]{0.45\textwidth}
    \centering
    \includegraphics[width=\textwidth]{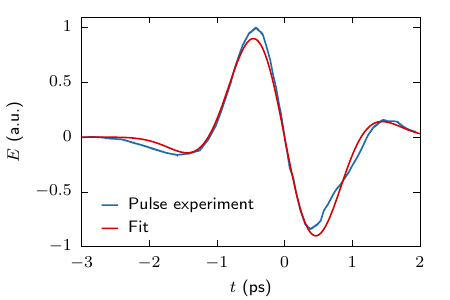}
    \caption{%
    \textbf{Comparison between experimental pulse
        and theoretical parametrization.}
    The electrical field of a single cycle THz quench pulse
    as used in experiment.
    The parameters of a fit with a Gaussian shape,
    as given in Equation~(20) of the main text,
    are given by $\hbar \omega_{\textrm{p}} = 1.33$\,meV
    and $\tau_{\textrm{p}} = 1.63$\,ps.
    }
    \label{fig:sfigure2}
\end{minipage}
\end{figure}

\section*{Supplementary Note 3: Analysis of realistic quench pulse}
In order to verify that our description of the experimental pulse is valid,
we performed measurements of the electric field
of a single cycle THz laser pulse.
From the theoretical parametrization in Equation~(20) of the main text
we deduce the field $E(t)$ and fit it to the experimental results
in Supplementary Figure~\ref{fig:sfigure2}.
The fit provides an excellent approximation to the pulse used in experiment.
In our simulations we rescaled the pulse width to $\tau_{\mathrm p} = 0.4$~ps
and the pulse frequency to $\hbar \omega_{\textrm{p}} = 3.0$~meV
to adjust it to the gap parameter used.

\begin{figure}[t]
\begin{minipage}[t]{0.45\textwidth}
    \centering
    \includegraphics[width=\textwidth]{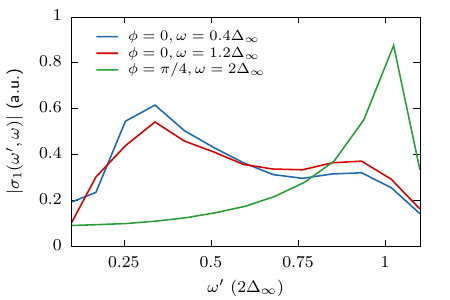}
\end{minipage}\hfill
\begin{minipage}[t]{0.45\textwidth}
    \centering
    \includegraphics[width=\textwidth]{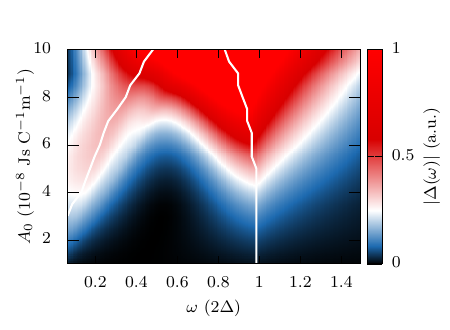}
\end{minipage}
\begin{minipage}[t]{0.45\textwidth}
    \caption{%
    \textbf{Fourier transform in the delay time $\Delta t$
    of the optical conductivity for a $d$-wave superconductor
    after a quench pulse.}
    Fourier spectrum $|\sigma_1(\omega',\omega)| = |\operatorname{FT} |\sigma_1(\Delta t,\omega)||$ of the real part of the optical conductivity
    for fixed frequency slices $\omega$ in Figure~4 of the main text.
    The red and the blue line display Higgs oscillations
    for a quench pulse with $\phi = 0$.
    The green curve corresponds to $\phi = \pi/4$.
    \label{fig:sfigure3}
    }
\end{minipage}\hfill
\begin{minipage}[t]{0.45\textwidth}
\caption{%
    \textbf{Fluence dependence of the gap dynamics for $\bm{\phi = 0}$}
    Fourier spectrum $|\Delta(\omega)| = |\operatorname{FT} |\Delta(t)||$ of the Higgs oscillations
    as a function of energy and laser fluence.
    The curves indicate the position of the $2\Delta$ peak
    and of the lower-lying peak.
    \label{fig:sfigure4}
    }
\end{minipage}
\end{figure}

\section*{Supplementary Note 4: Fourier Transform of the optical conductivity}
The optical conductivity shown in Figure~4 of the main text
displays Higgs oscillations as function of the time delay $\Delta t$
between quench and probe pulse.
In Supplementary Figure~\ref{fig:sfigure3}
we analyze the Fourier spectrum of the oscillations,
to verify that they correspond to the Higgs oscillations of the gap.
To achieve this, we fix the frequency $\omega$ of the optical conductivity
and Fourier transform the optical conductivity
as function of the time delay $\Delta t$.
We denote the corresponding frequency axis with $\omega'$
to distinguish it from the frequency $\omega$.
Clearly the optical conductivity shows the same peak structure
as the Higgs oscillations,
depending strongly on the incident angle $\phi$.
Thus the optical conductivity serves as an effective tool
to measure Higgs oscillations.

\section*{Supplementary Note 5: Fluence dependence of the Higgs mode}

Besides the angular dependence of the Higgs mode
in simulations with realistic quench pulses,
it also depends on the intensity of the quench pulse.
In Supplementary Figure~\ref{fig:sfigure4}
the Fourier transform of $\Delta(t)$
is shown for various quench pulse intensities.
The $2\Delta$ Higgs mode depends
on the asymptotic value of the order parameter.
With increasing intensity, more and more quasi-particles are excited
and thereby the order parameter is reduced.
Due to this mechanism, the frequency of the symmetric Higgs mode decreases
with increasing intensity.
This behavior is similar to the $s$-wave superconductor \cite{Yuz06}.
In contrast, the frequency of the second mode at lower energy
is increased for higher fluence until it merges with the $2\Delta$ Higgs mode.
An increase in fluence corresponds to an increase in the quench strength.
As discussed,
the second mode depends strongly on the difference
$f(\varphi)^2 - f'(\varphi)^2$,
which increases for higher intensity of the laser.
This leads to the shift to higher energy for stronger laser fluence.

\section*{Supplementary Note 6: Classification of condensate oscillations}
\label{sec.classification}
To classify all possible condensate oscillations,
we quench all fundamental gap symmetries with all fundamental quenches
allowed by the $D_{4\mathrm{h}}$ point group
from Supplementary Table~\ref{tab.basis_functions}.
In Supplementary Figure~\ref{fig:sfigure5},
we show the quenched symmetry functions for
all combinations of gap and quench symmetry.
After the quench, we perform a time-evolution,
where we extract for each step of the time-evolution
the symmetry of the condensate
$\braket{c_{-\vk\downarrow}c_{\vk\uparrow}}$
for each step of the time-evolution and classify its oscillation.
According to Equation~(10) of the main text,
the expectation value in equilibrium reads
\begin{align}
    \braket{c_{-\vk\downarrow}c_{\vk\uparrow}} = \frac{\Delta f_\vk}{2E_\vk}\,.
\end{align}
Using our assumption that $\epsilon_\vk = \epsilon(|\vk|)$
and $f_\vk = f(\varphi)$,
for $k$-values far away from $k_{\mathrm F}$ it follows
\begin{align}
    \braket{c_{-\vk\downarrow}c_{\vk\uparrow}}
        \approx \frac{\Delta f_\vk}{2\epsilon_\vk} \propto f(\varphi)\,.
\end{align}
We therefore use $\braket{c_{-\vk\downarrow}c_{\vk\uparrow}}$ at a $k$-value
far from $k_\mathrm{F}$ as an approximation
to trace the dynamics of the symmetry of the condensate.
An animation of the condensate dynamics
can be found in the Supplementary Movie 1.
We name the oscillations according to its symmetry.
Finally, we calculate the gap oscillation and its Fourier spectrum
to obtain the collective Higgs modes
resulting from the condensate oscillations.

\begin{table}[t]
\setlength{\tabcolsep}{9pt}
\begin{tabular*}{0.5\linewidth}{lll}
    \toprule
    \multirow{2}{*}{\shortstack{Irreducible\\representation}} &
    \multirow{2}{*}{Basis function} &
    \multirow{2}{*}{\shortstack{Basis function\\polar coordinates}} \\\\
    \toprule
    $A_{1\mathrm{g}}$ & $1$ & $1$\\
    \midrule
    $A_{2\mathrm{g}}$ & $xy(x^2-y^2)$ & $\sin(4\varphi)$\\
    \midrule
    $B_{1\mathrm{g}}$ & $x^2-y^2$ & $\cos(2\varphi)$ \\
    \midrule
    $B_{2\mathrm{g}}$ & $xy$ & $\sin(2\varphi)$ \\
    \bottomrule
    \end{tabular*}
    \caption{\textbf{Basis functions for $D_{4\mathrm{h}}$ point group}
        Even parity basis functions for the $D_{4\mathrm{h}}$ point group
        in Cartesian coordinates and
        as function of the azimuthal angle $\varphi$ in polar coordinates
        \cite{RevModPhys.63.239}.}
    \label{tab.basis_functions}
\end{table}

\begin{figure*}[t]
    \centering
    \includegraphics[width=\textwidth]{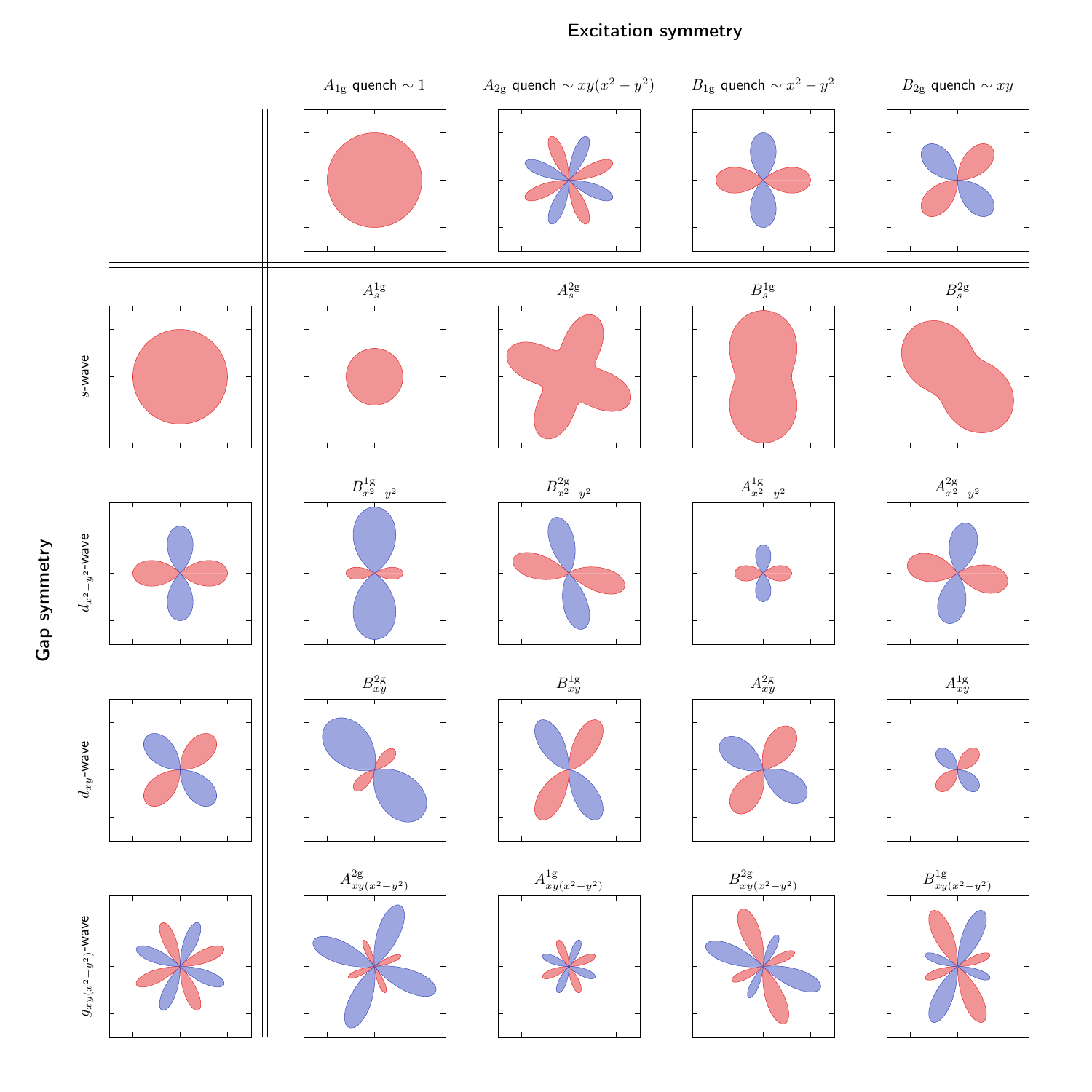}
    \caption{%
    \textbf{Oscillation symmetries of the $D_{4\mathrm{h}}$ point group}
    Oscillation symmetries of all fundamental gap symmetries allowed
    by the $D_{4\mathrm{h}}$ point group.
    The left column depicts the four symmetries
    $s$-, $d_{x^2-y^2}$-, $d_{xy}$- and $g_{xy(x^2-y^2)}$-wave
    i.e. the gap symmetry function $f_\vk$.
    The top row shows possible excitation
    or quench symmetries $f_\vk^\mathrm{q}$.
    Each cell in the graphics grid corresponds to the deformation of the
    gap symmetry with the respective quench symmetry, i.e. $f'_\vk = f_\vk + \delta f_\vk^{\mathrm q}$ with an exemplary value of $\delta = -0.4$.
    The labels indicate the induced oscillations.
    An animation of the oscillations can be found in the Supplementary Movie 1.
    }
    \label{fig:sfigure5}
\end{figure*}